\documentclass[twocolumn,showpacs,preprintnumbers,prb,amsmath,amssymb,floatfix]{revtex4}
\usepackage{graphicx}% Include figure files
\usepackage{epsfig} %for old style fig. inputs
\usepackage{dcolumn}% Align table columns on decimal point
\usepackage{bm}% bold math

\begin{document}
\newcommand{\rth}{$R_{th}$ }
\newcommand{\rthy}{$R_{th}$}
\newcommand{\cll}{$c_{11}$ }
\newcommand{\clly}{$c_{11}$}
\newcommand{\cuu}{$c_{66}$ }
\newcommand{\cuuy}{$c_{66}$}
\newcommand{\cms}{cm$^{2}$ }
\newcommand{\cmsy}{cm$^{2}$}
\newcommand{\cmc}{cm$^{3}$ }
\newcommand{\cmcy}{cm$^{3}$}
\newcommand{\ab}{$\sim$ }
\newcommand{\aby}{$\sim$}
\newcommand{\tp}{$T'$ }
\newcommand{\tpy}{$T'$}
\newcommand{\tph}{$T_{p}$ }
\newcommand{\tphy}{$T_{p}$}
\newcommand{\too}{$T_{0}$ }
\newcommand{\tooy}{$T_{0}$}
\newcommand{\tauep}{$\tau_{ep}$ }
\newcommand{\tauepy}{$\tau_{ep}$}
\newcommand{\tauee}{$\tau_{ee}$ }
\newcommand{\taueey}{$\tau_{ee}$}
\newcommand{\jphi}{$j_{\phi}$ }
\newcommand{\jphiy}{$j_{\phi}$}
\newcommand{\tc}{$T_{c}$ }
\newcommand{\tcy}{$T_{c}$}
\newcommand{\hcl}{$H_{c1}$ }
\newcommand{\hcly}{$H_{c1}$}
\newcommand{\ef}{$E_{F}$ }
\newcommand{\efy}{$E_{F}$}
\newcommand{\estar}{$E^{*}$ }
\newcommand{\estary}{$E^{*}$}
\newcommand{\htc}{high-temperature superconductors } 
\newcommand{\htcy}{high-temperature superconductors}
\newcommand{\et}{{\it et al. }}
\newcommand{\ety}{{\it et al.}}
\newcommand{\be}{\begin{equation} }
\newcommand{\ene}{\end{equation}}
\newcommand{\hh}{$H$ }
\newcommand{\hhy}{$H$}
\newcommand{\hc}{$H_{c}$ }
\newcommand{\hcy}{$H_{c}$}
\newcommand{\ho}{$H_{0}$ }
\newcommand{\hoy}{$H_{0}$}
\newcommand{\jc}{$j_{c}$ }
\newcommand{\jcy}{$j_{c}$}
\newcommand{\sg}{superconducting }
\newcommand{\sgy}{superconducting}
\newcommand{\ssc}{superconductor }
\newcommand{\sscy}{superconductor}
\newcommand{\hcu}{$H_{c2}$ }
\newcommand{\hcuy}{$H_{c2}$}
\newcommand{\rfff}{$\rho_{f}$ }
\newcommand{\rfffy}{$\rho_{f}$}
\newcommand{\hcut}{$H_{c2}(T)$ }
\newcommand{\hcuty}{$H_{c2}(T)$}
\newcommand{\jd}{$j_{d}$ }
\newcommand{\jdy}{$j_{d}$}
\newcommand{\id}{$I_{d}$ }
\newcommand{\idy}{$I_{d}$}
\newcommand{\ybco}{Y$_{1}$Ba$_{2}$Cu$_{3}$O$_{7}$ }
\newcommand{\ybcoy}{Y$_{1}$Ba$_{2}$Cu$_{3}$O$_{7}$}
\newcommand{\lsco}{La$_{2-x}$Sr$_{x}$CuO$_{4}$ }
\newcommand{\lscoy}{La$_{2-x}$Sr$_{x}$CuO$_{4}$}
\newcommand{\mgb}{MgB$_{2}$ }
\newcommand{\mgby}{MgB$_{2}$}
\newcommand{\de}{$\delta \epsilon$ }
\newcommand{\dey}{$\delta \epsilon$}
\newcommand{\nq}{$n_{q}$ }
\newcommand{\nqy}{$n_{q}$}
\newcommand{\rrhon}{$\rho_{n}$ }
\newcommand{\rrhony}{$\rho_{n}$}
\newcommand{\rrho}{{$\rho$} }
\newcommand{\rrhoy}{{$\rho$}}
\newcommand{\qp}{quasiparticle }
\newcommand{\qpy}{quasiparticle}
\newcommand{\qps}{quasiparticles }
\newcommand{\qpsy}{quasiparticles}
\newcommand{\bib}{\bibitem}
\newcommand{\ib}{{\em ibid. }}
\newcommand{\taue}{$\tau_{\epsilon}$ }
\newcommand{\tauey}{$\tau_{\epsilon}$}
\newcommand{\vstary}{$v^{*}$}
\newcommand{\vstar}{$v^{*}$ }
\newcommand{\tstar}{$T^{*}$ }
\newcommand{\tstary}{$T^{*}$}
\newcommand{\rhostar}{$\rho^{*}$ }
\newcommand{\rhostary}{$\rho^{*}$}
\newcommand{\vinf}{$v_{\infty}$ }
\newcommand{\vinfy}{$v_{\infty}$}
\newcommand{\fd}{$F_{d}$ }
\newcommand{\fdy}{$F_{d}$}
\newcommand{\fe}{$F_{e}$ }
\newcommand{\fey}{$F_{e}$}
\newcommand{\fl}{$F_{L}$ }
\newcommand{\fly}{$F_{L}$}
\newcommand{\jstar}{$j^{*}$ }
\newcommand{\jstary}{$j^{*}$}
\newcommand{\je}{$j(E)$ }
\newcommand{\jey}{$j(E)$}
\newcommand{\vphi}{$v_{\phi}$ }
\newcommand{\vphiy}{$v_{\phi}$}
\newcommand{\blo}{$B_{1}$ }
\newcommand{\bloy}{$B_{1}$}
\newcommand{\bhi}{$B_{\infty}$ }
\newcommand{\bhiy}{$B_{\infty}$}
\newcommand{\vlo}{$v_{1}$ }
\newcommand{\vloy}{$v_{1}$}
\newcommand{\bo}{$B_{o}$ }
\newcommand{\boy}{$B_{o}$}
\newcommand{\eo}{$E_{o}$ }
\newcommand{\eoy}{$E_{o}$}

\preprint{Published in Phys. Rev. B (Rapid Comm.) {\bf 68}, 100503 (2003).}

\title{Critical flux pinning and enhanced upper-critical-field in magnesium
diboride films}

\author{Milind N. Kunchur} 
 \homepage{http://www.physics.sc.edu/kunchur}
 \email{kunchur@sc.edu}
\author{Cheng Wu} 
\author{Daniel H. Arcos} 
\author{Boris I. Ivlev} 
\altaffiliation[Also at ]{Instituto de F\'{\i}sica, Universidad Autonoma
de San Luis Potos\'{\i}, S.L.P. 78000 Mexico}
%%\author{}%
%% \email{Second.Author@institution.edu}
\affiliation{Department of Physics and Astronomy\\
University of South Carolina, Columbia, SC 29208}

\author{Eun-Mi Choi}
\author{Kijoon H.P. Kim}
\author{W. N. Kang}
\author{Sung-Ik Lee}

%% \homepage{http://www.Second.institution.edu/~Charlie.Author}
\affiliation{National Creative Research Initiative Center for
Superconductivity\\and
Department of Physics\\
Pohang University of Science and Technology\\Pohang 790-784, Republic of Korea}
%%Second institution and/or address\\
%%This line break forced% with \\
%%}%

\date{\today}% It is always \today, today,

\begin{abstract}
We have conducted pulsed transport measurements on $c$-axis oriented 
magnesium diboride films over the entire relevant ranges of magnetic
field $0 \alt H \alt H_{c2}$ (where \hcu is the upper critical field) 
and current density $0 \alt j \alt j_{d}$ (where 
\jd is the depairing current density).
 The intrinsic disorder of the films combined with the large coherence
length and three-dimensionality, compared to cuprate
superconductors, results in a six-fold 
enhancement of \hcu and raises the depinning current density \jc to
within an order of magnitude of \jdy . 
The current-voltage response is highly non-linear at all fields,
resulting from a combination of depinning and pair-breaking, and has no
trace of an Ohmic free-flux-flow regime. 
\end{abstract}

%http://www.aip.org/pacs/pacs03/pacs0370.html
%Valid PACS appear here}% PACS, the Physics and Astronomy
%Suggested keywords}%Use showkeys class option if keyword
\maketitle

\section{\label{sec:level1}Introduction}

Magnesium diboride (\mgby ) recently made an impact as a promising new 
superconductor with a surprisingly high critical temperature \tc 
for a simple binary compound. Besides the temperature $T$, the other
principal parameters that define the operational space of a
superconductor are magnetic field $H$ and current density $j$.
Superconductivity perishes above \hcuy $(T,j)$ and \jdy $(T,H)$. For most
practical applications it is not sufficient for the system to be merely
in a superconducting state (presence of a 
finite order parameter amplitude) but for the system to be in a 
dissipationless state (constancy of order-parameter phase difference
across sample length). Thus from a practical standpoint, 
the conventional critical current density \jc is limited by the 
depinning of vortices and their consequent motion. In the high-\tc
cuprates, the large \tc is accompanied by 
high values \cite{pair} of \jd ($>10^{8}$ A/\cmsy ). However the
layered structure and small coherence length $\xi$ lead to very weak
vortex pinning.
Thus between \jc and \jdy , there can be a broad dissipative regime \cite{mplb}.

\mgb has an intermediate \tc and low values of \jd and \hcuy . 
\jd $\sim 10^{7}$ A/\cms at low temperatures and, 
in single crystals, \hcu \ab 3 T parallel to the c-axis \cite{ott}.
In the films studied here, the
intrinsic disorder makes two striking improvements. First of all \hcu is
enhanced six-fold. Secondly, pinning is enhanced to the point of
raising \jc to the same order of magnitude as \jdy . These observations are
consistent with the reduction of $\xi$ with disorder. Finally  \mgby 's
normal-state resistivity \rrhon is lower than that of a typical cuprate 
superconductor (e.g., \ybcoy ) by two orders
of magnitude. This increases the vortex viscosity by the same factor, so
that even when the vortices are depinned, they flow at very low
velocities causing insignificant dissipation until \jc becomes
comparable to \jdy . All these factors lead to an extremely steep
non-linear current-voltage ($IV$) curve with a complete absence of an
Ohmic regime characteristic of free flux flow \cite{obs}.  

%{\bf Experimental details:}
\section{Experimental details}
\begin{figure}[h] 
\begin{center}
%%\vspace{1.5cm}
	\leavevmode
	\epsfxsize=0.9\hsize
	%% include at end to save time! \epsfbox{SamplePulses.eps} %S-RT.eps}
	\epsfbox{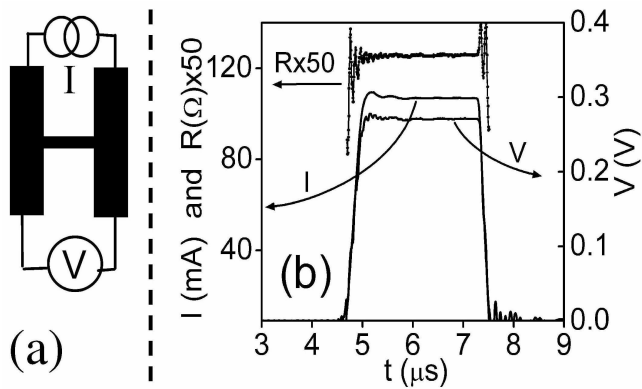} %SamplePulses.eps} %S-RT.eps}
%from 0708g1-and-0705g1.OPJ/pulses.jpg/pulses.eps, 
%and circuit.jpg/circuit.eps 
%combined within Sample-shape-and-pulses.ppt
\end{center}
\vspace{-2em}
\caption{\label{exptl}{\em 
(a) Sample geometry used for resistance measurement. At low
values of $j$ the wide lead areas add a constant resistance of about 
15\% of the total value. At high $j$ this contribution is frozen out.
(b) Pulse waveforms under worst-case conditions ($j = 9.7$ MA/\cmsy, 
$E = 83$ V/cm, and  $p = jE = 803$ MW/\cmc on the plateaus). 
 The resistance rises to (90\% of) its final value in about 50
ns from the (10\%) onset of $I$.}}
\end{figure}

The samples are 400 nm thick films of MgB$_{2}$ 
fabricated using a two-step method whose details are described 
elsewhere\cite{sampleprep,sampleprep2}. An amorphous boron film was 
deposited on a (1\={1}02) Al$_{2}$O$_{3}$ substrate at room temperature by 
pulsed-laser ablation. The boron film
was put into a Nb tube with high-purity Mg metal (99.9\%)
and the Nb tube was then sealed using an arc furnace in an
argon atmosphere. Finally, the tube was heated to 900$^{\circ}$ 
C for 30 min. in an evacuated quartz ampoule 
sealed under high vacuum. X-ray diffraction indicates 
a highly c-axis-oriented crystal structure normal to 
the substrate surface with no impurity phases.
The films were photolithographically patterned down to narrow bridges.
In this paper we show data on three bridges, labelled S, M, and L (for
small, medium, and large) with lateral dimensions 
2.8 x 33, 3.0 x 61, and 9.7
x 172 $\mu$m$^{2}$ respectively. The lateral
dimensions are uncertain by $\pm 0.7 \mu$m and the thickness by $\pm 50$ nm.
%more exactly: 2.77x32.6; 3.01x60.5, 9.73x172
Fig.~\ref{exptl}(a) shows the sample geometry. 
The horizontal sections of the current leads add a $\sim 15$ \% series resistance to the resistance of the actual bridge,
which enters in measurements made at very small $j$. At high $j$ in the
mixed state, the extra contribution is frozen out because the current is
 spread out in these wider areas, diluting its density, so that
one observes mainly the resistance of the bridge (A more intricate
contact geometry was prohibited because of the extreme difficulty we
faced in etching \mgby .).  
%\begin{figure}[ht]

\begin{figure}[ht] 
\begin{center}
%%\vspace{1.5cm}
	\leavevmode
	\epsfxsize=0.7\hsize
	\epsfbox{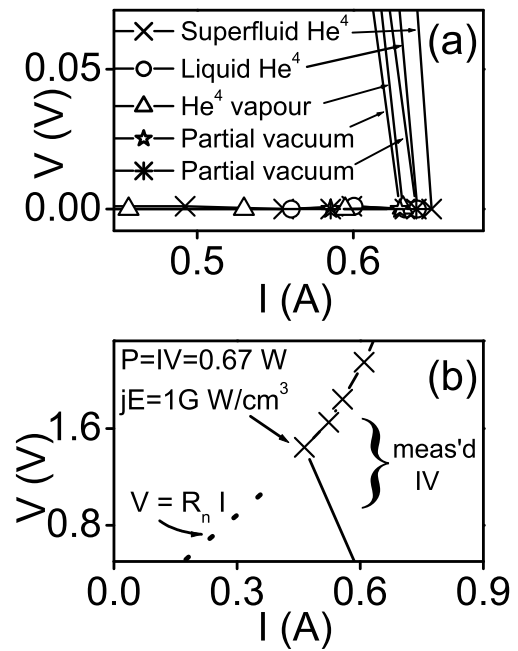} %FigEnvs.eps} %S-RT.eps}
%%from mgflow3.tex which combines PBfigMivs.eps and PBfigEnvs.eps
\end{center}
\vspace{-2em}
\caption{\label{pbcurrent}{\em Zero-field IV 
curves for sample L show an abrupt switch to a state of full 
 normal-state resistance, $R_{n}$, by currents of
pair-breaking magnitudes. (a) The switch to
 a dissipative state is not significantly
influenced by the thermal environment. (b) Top portion of the curve in
superfluid He$^{4}$. The resistive portion of the measured curve
extrapolates to the $V=R_{n}I$ dashed line.}}
\end{figure}

The non-linear electrical transport measurements were made using a
pulsed signal source
with pulse durations ranging from 0.1 to 4 $\mu$s and a duty cycle of about 1
ppm. At the other extreme, a conventional continuous DC method at a 
very low current ($I= 1.4 \mu$A) was employed for the
resistive traces used to determine \hcuy . 
Fig.~\ref{exptl}(b) shows pulse waveforms under the especially severe
conditions of $j = 9.7$ MA/\cmsy, $E = 83$ V/cm, and  $p = jE = 803$
MW/\cmcy . The resistance rises  to 90\% of 
its final value in about 50 ns from the 10\% onset of $I$. 
From a knowledge of the thermal conductivities
%, $\sigma$, 
and specific heat capacities of the film and substrate materials, and their
mutual thermal boundary resistance, one can calculate the total thermal
resistance  $R_{th}$ for any pulse duration \cite{mplb,gupta}. Also if $R(T)$
has enough variation, the film's own resistance can be used as a
thermometer to measure \rthy . For films of  \ybco (YBCO) on
LaAlO$_{3}$, which were used for most of our previous work, we found 
$R_{th} \sim$ 1--10 nK.cm$^{3}$/W at
microsecond timescales \cite{unstable,metal,mplb}. In the
case of our \mgb films, we expect \rth to be smaller because of
sapphire's very high conductivity. However the five parameters required
to calculate \rth from first principles 
are not all known for this film-substrate combination and
\mgb has a very flat $R(T)$ below 50 K, so one can't measure \rth as was
done for YBCO. We can, however, obtain an upper bound on \rth 
in the following way: Fig.~\ref{pbcurrent}(a) shows 
$IV$ curves for sample L in zero field 
(This is the largest sample with the lowest surface-to-volume
ratio, so that it represents the worst case thermal scenario.). 
The curves were measured with the sample in different thermal
environments. Above some threshold current $I_{d}$ \ab 650 mA, the system
abruptly switches into the normal state. The value of $I_{d}$ 
is not sensitive to the thermal environment contacting  
the exposed surface  of the film, confirming 
that the highly conductive sapphire,
together with the greatly reduced heat input during the short pulse,
prevents the film's temperature from rising significantly (It has
been shown by Stoll et al. \cite{stoll} that if there is sample heating,
the thermal environment makes a significant difference because it will
provide an additional path for the heat to flow through.). 
Fig.~\ref{pbcurrent}(b) shows the top portion of one of the IV 
curves. The resistance jumps directly 
to the full normal-state value (The arrow indicates the first data point with 
non-zero resistance; the dashed line corresponds to $V = R_{n}I$.).
 It is argued elsewhere
\cite{mgb2pb} that this jump to the normal state occurs due to
pairbreaking by the current \footnote{The value of the current density
at which this jump occurs, \aby $2 \times 10^{7}$ A/\cmsy , is roughly 
comparable to the theoretical estimate of 
$j_{d}(0)=cH_{c}(0)/[3\sqrt{6}\pi \lambda(0)]
\sim 6 \times 10^{7}$ A/\cmsy .}. At the point the system is just driven
normal, the power density reaches
 $p=jE=1.01$ GW/\cmc  (arrow in Fig.~\ref{pbcurrent}(b)). This sets
%Average current for sample M is ~ 0.02A => j=1.67A/cms
%R ~ Rn/2, so p = 19 MW/cmc . 
a gross upperbound of $R_{th} \sim 7$ nK.cm$^{3}$/W. Note that
the main bottle neck of heat conduction 
is the film-substrate boundary resistance which is not strongly temperature
dependent \cite{gupta}.  In the present work, typical $p$ values are
two orders of magnitude lower than the critical 1 GW/\cmc and so we expect 
the temperature rise to be a small fraction (\aby 1 \%) of \tcy .

The magnetic field is applied normal to the film (parallel to the {\em
c} axis) and the self field of the current ($< 50$ G) 
is much lower than the applied fields used in this work. 
Further details of the
measurement techniques have been published in a previous review article
\cite{mplb} and other recent papers \cite{metal,unstable}. 
%\begin{figure}[ht] 

\section{Results and analysis} %\label{sec:level1}}
\begin{figure}[ht] 
\begin{center}
%%\vspace{1.5cm}
	\leavevmode
	\epsfxsize=0.8\hsize
	\epsfbox{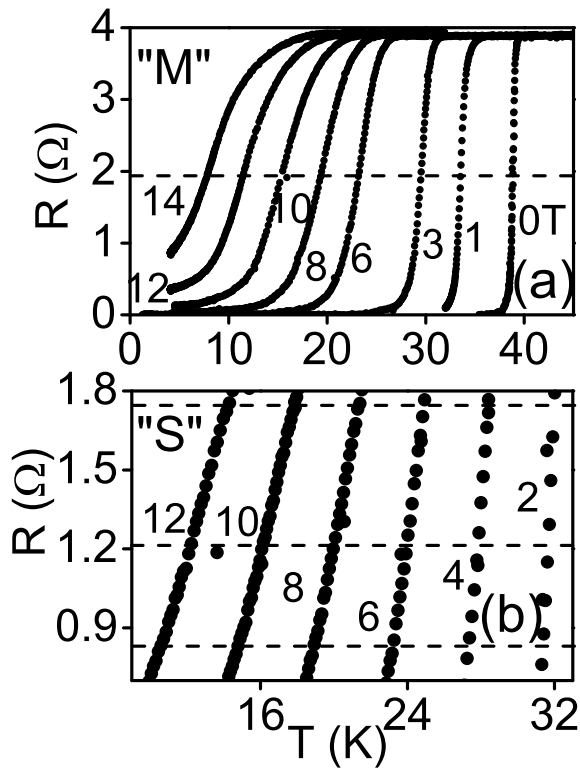} %RT-B-curves.eps} %S-RT.eps}
%%from 081002b2.opj (small) and 080702c1.opj (medium)
\end{center}
\caption{\label{rtcurves}{\em Resistive transitions of two \mgb bridges at
indicated flux densities. I=1.4 $\mu$A. (a) Sample M;
% at flux densities $B= 0, 1, 3, 6, 8, 10, 12, 14$ T from right to left.
midpoint \tcy 's lie on the 
dashed line through $R=R_{n}/2=1.95 \Omega$.
(b) Similar set for 
sample S. Dashed lines run through $R=R_{n}/2=1.22 \Omega$, and also
through the 
lower and higher resistive criteria $R=0.84 \Omega$ and $R=1.76 \Omega$.}}
%\end{center}
\end{figure}

\begin{figure}[ht] 
\begin{center}
%%\vspace{1.5cm}
	\leavevmode
	\epsfxsize=0.98\hsize
	\epsfbox{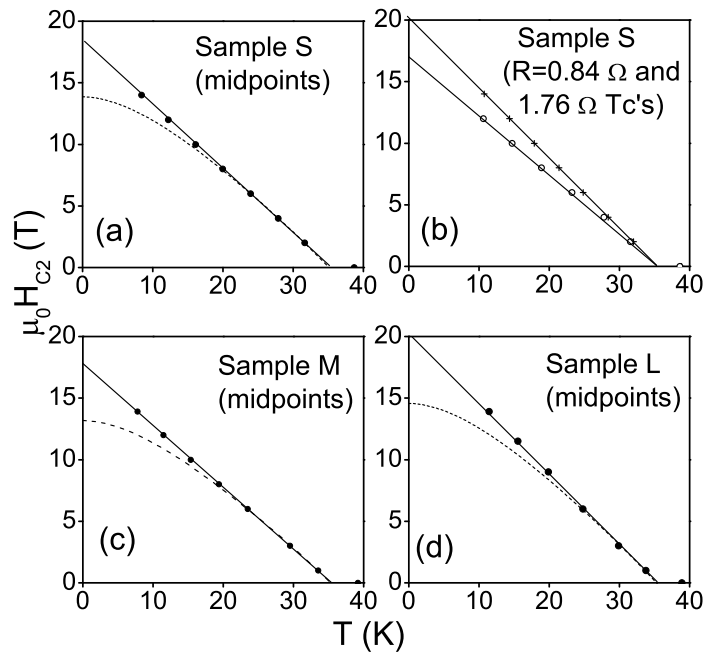} %tcb.eps} %S-RT.eps}
%%from 081002b2.opj (main one) containing graphs from 
%%081202a2.opj, 080702c1.opj
\end{center}
\caption{\label{tcb}{\em \hcu versus $B$ for three samples. The solid
straight lines are linear fits to the data. The dashed lines correspond
to the WHH function. Panel (b) shows \hcu for sample S defined at two
other criteria. The linear behavior of \hcuy $(T)$ 
continues to hold 
for the R=0.84 $\Omega$ and R=1.76 $\Omega$ criteria.}}
%\end{center}
\end{figure}

%{\bf Results and analysis:}
Fig.~\ref{rtcurves} shows the resistive transitions at a low
continuous current of $I= 1.4 \mu$A in different fixed
magnetic fields.  Panel (a) shows the full curves for sample M, whereas for
sample S we show the central region of the transition in panel (b). Both
the full and central views look very similar for all three samples. 
In the central region,
the shifts are approximately parallel, reducing the ambiguity of $H_{c2}(T)$ 
defined by the transition-midpoint criterion. Fig.~\ref{tcb} shows those 
$H_{c2}(T)$ values. $H_{c2}(T)$ has an unusual linear dependence
that departs noticeably from the dashed line of the WHH 
(Werthamer-Helfand-Hohenberg) function \cite{whh}. This observation is
consistent with that of a dirty two-gap superconductor \cite{gurevich}. 
Panel (b) shows $H_{c2}(T)$ for sample S defined by other criteria (see
Fig.~\ref{rtcurves}(b) above). Note that the linear dependence is
preserved and the slopes (and hence the inferred zero-$T$ values of \hcuy )
do not change significantly with the choice of criteria. We infer a value of $17
\pm 3$ T for \hcuy $(0)$, which is six times higher than the
value of 3 T found in single crystals \cite{ott}. This can be
understood in view of the much
higher resistivity and hence shorter coherence length $\xi$: For a dirty
superconductor $\xi \approx \sqrt{\xi_{0} l}$, where $\xi_{0}$ is the 
clean-limit value of $\xi$ and $l$ is the impurity scattering mean-free-path.
Since $\rho \propto l$ and \hcu $\propto 1/\xi^{2}$,  
we expect  \hcu $\propto \rho$. The normal resistivity of our film
samples (14 $\mu \Omega$-cm) is about seven times higher than 
the value (2 $\mu \Omega$-cm) of the single crystals of Ref.~\cite{ott}.
Note that this
\hcu enhancement occurs naturally in the as-prepared films without any special
effort to introduce pinning sites. The microscopic cause of the enhanced
scattering leading to the higher \hcu in these samples is presently not clear.
%\begin{figure}[ht] 

We now turn to the in-field IV curves to investigate the nature of flux
motion.  In a system with weak flux pinning, the
resistance goes through alternate regimes of Ohmic ($V \propto I$) 
and non-Ohmic  behavior. At very low driving
forces (low $j$) there can be observable resistance due to thermally
activated flux flow (TAFF) or flux creep. Then one encounters a
non-linear response as current-driven depinning sets in; in effect the
number of mobile vortices is rising. This is incipient flux flow. 
 At sufficiently larger $j$, the vortex motion 
is effectively free from the influence of pinning and the response
becomes Ohmic again. We previously introduced the term {\em free flux
flow} (FFF) for this linear regime \cite{obs}. Here the dissipation and
resistivity should correspond to the canonical $\rho_{f} \sim \rho_{n}
B/$\hcu Bardeen-Stephen expression\footnote{Sometimes 
large departures can occur for exceptional situations such as
superclean systems and narrow vortex cores where the internal
energy-level spacing exceeds their widths.}.  
 As one goes beyond FFF, non-linearity can set in because of
heating of the electron gas \cite{unstable} or changes in the electron
distribution function \cite{lo}. Finally at yet higher currents, 
pair-breaking destroys superconductivity and drives the system normal.
Here the resistance again ceases to change with current, being
characteristic of the normal state,  and so the
response becomes Ohmic one more time. These stages of dissipation have
been described in our previous review article \cite{mplb}. In \ybcoy
, the depinning critical current is sufficiently weak compared to the
pair-breaking value that all of the regimes can be observed. 
%\begin{figure}[ht] 

\begin{figure}[ht] 
\begin{center}
%%%\vspace{1.5cm}
	\leavevmode
	\epsfxsize=0.7\hsize
	\epsfbox{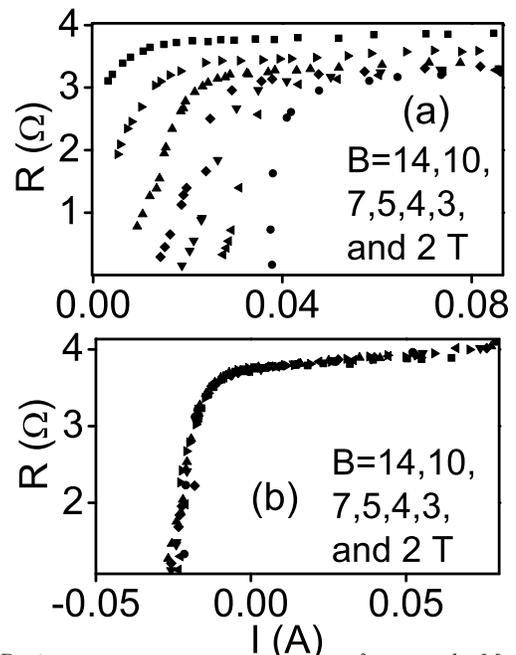} %SampleCcurves.eps} %S-RT.eps}
%%	\epsfbox{SampleCcurvesColor.eps} %S-RT.eps}
%%from mgb2flowFFcurves.opj which contains .ogg files 
%%from the original origin workbook/s "flux flow for  sample C.opj", etc.
\end{center}
\vspace{-3em}
\caption{\label{mgbflow}{\em Resistance versus current curves for sample
M. Flux densities are indicated from left to right. The
sample was immersed in superfluid helium and $T= 1.5$K for all data. 
 (a) Raw data. (b) Same data linearly shifted so as to collapse onto a 
single common curve.}}
%\end{center}
\end{figure}

In \mgb the situation is very different. Fig.~\ref{mgbflow}(a) 
shows the R(I) curves of the present \mgb samples. After the onset of
dissipation, the resistance quickly rises to the full normal-state
value. It should be noted that the plateaus do not correspond to FFF but
to the normal state. Accordingly the resistance value changes very little with
the applied $B$, especially for the curves at lower fields\footnote{The
slight shift in plateau resistance at the highest 
fields can be understood in terms of spreading of resistance outside the
bridge area and into the current-lead areas as explained in the experimental
section. Fields approaching \hcu start driving the
whole film normal at relatively low currents so that the resistance of
the wider current-lead areas is not frozen out. This causes the
normal-state plateau to rise slightly at the highest fields.}.
 The overall shapes of
the curves are almost independent of field. When the curves
of the previous panels are shifted vertically and horizontally by
constant amounts they can be made to overlap as shown in Fig.~\ref{mgbflow}(b).

\begin{figure}[h] 
%\label{scaled}
\begin{center}
%%\vspace{1.5cm}
	\leavevmode
	\epsfxsize=0.8\hsize
	\epsfbox{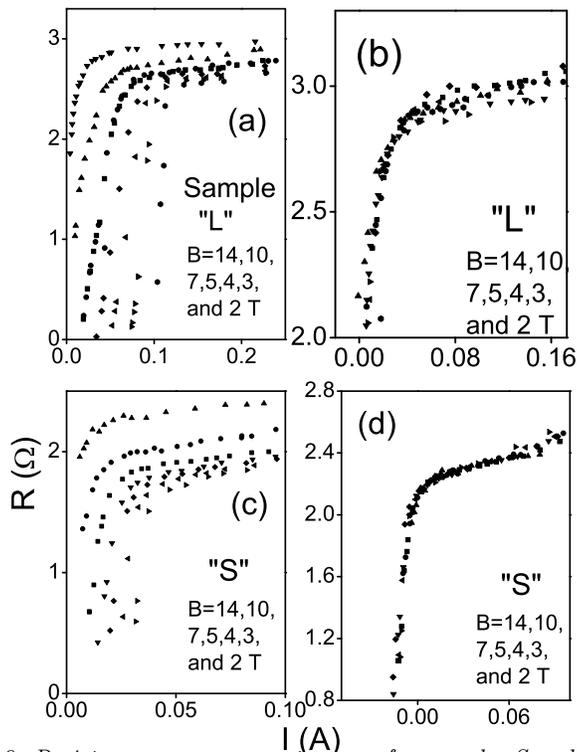} %SamplesBCcurves.eps} %S-RT.eps}
%	%\epsfbox{SamplesBCcurvesColor.eps} %S-RT.eps}
%%from mgb2flowFFcurves.opj which contains .ogg files 
%%from the original origin workbook/s "flux flow for  sample C.opj", etc.
%%\end{center}
~\vspace{-2em}~\caption{\label{othersamples}{\em Resistance 
versus current curves for samples S and L. Flux 
densities are indicated from left to right.
The sample was immersed in superfluid helium and $T= 1.5$K for all data. 
Panels in the left column show raw data and panels in the right column show
linearly shifted data.}}
\end{center}
\end{figure}
Fig.~\ref{othersamples} shows the raw and shifted curves for 
the other two samples. 

These steep IV characteristics in \mgb are similar to those for 
niobium alloys but are markedly different from those in the cuprates. 
The latter can be understood in terms of \mgby 's lower
temperature scale and much greater pinning (due to its higher 
isotropy and ten times larger vortex cross section). 
%$ = 5  instead of 1.6 nm). 
Hence thermal activation and 
current driven depinning are deferred until $j$ becomes almost
comparable to \jdy , causing the response to become steeply non-linear, rising
from very little dissipation to the full normal
state within a rather narrow range of currents. The collapse in
Figs.~\ref{othersamples} and \ref{mgbflow}
 seems to suggest that the rise in resistance stems 
mainly from the intrinsic mechanisms of current- and field-induced 
pair-breaking, rather than from current driven depinning, even when $B$
approaches \hcuy . This makes \mgby 's mixed-state response unique 
among studies of type-II superconductors. From the standpoint of
applications, this implies that
the practically useable current densities are much higher than might be
inferred from the \jd of the material, since substantial 
flux dissipation is deferred until \jc becomes comparable to \jdy .

\section{Conclusions} We have investigated the 
low-temperature ($T \ll T_{c}$) in-field transport behavior of
\mgb and present the first
measurement of the full dissipation 
curves (i.e., $0 \leq j \alt j_{d}$ and $0 \leq R_{T=0} \alt R_{n}$) 
for this system. \mgb films made by the two-step laser-ablation
process have an intrinsic pinning of a critical magnitude, such that the 
principle dissipation and resulting IV curves
arise mainly from intrinsic mechanisms such as pair-breaking. 
The onset of dissipation is within an order of magnitude of the
pair-breaking current, even at flux densities of a few teslas---the 
resistance rising quickly to the full normal-state value as the
current is increased beyond \jcy . The same disorder that enhances pinning, also
enhances \hcu and qualitatively changes its temperature dependence because of
the two-band nature of the superconductivity in this material. As
explained by the theory of Gurevich, \hcuy$(0)$ extrapolates to a higher
value than would be expected for WHH behavior, and the slope, 
$d H_{c2}/d t$, is relatively constant. 
Our measurements of \hcu are consistent with this predicted unusual
dependence. Both the enhancement in \hcu and the critically pinned flux 
make such \mgb films more promising for applications, besides providing
a cleaner and more intrinsic view of a superconductor's
current-voltage response in the mixed state. 

\vspace{1em}

\section{Acknowledgements} 

The authors acknowledge useful discussions with J. M.
Knight and A. Gurevich. This work was supported by the U. S. Department of
Energy through grant number DE-FG02-99ER45763 and 
by the Ministry of Science and
Technology of Korea through the Creative Research Initiative Program.

\end{document}